# Plasma burn-through simulations using the DYON code and predictions for ITER

Hyun-Tae Kim[1,2], A.C.C. Sips[1,3], P.C. de Vries[4], and JET-EFDA Contributors*

JET-EFDA Culham Science Centre, Abingdon, OX14 3DB, UK.

E-mail: hyun.kim09@imperial.ac.uk

1) Department of Physics, Imperial College London, SW7 2AZ, London, UK.
2) EURATOM/CCFE Association, Culham Science Centre, OX14 3DB Abingdon, UK.
3) EFDA-CSU, Culham Science Centre, OX14 3DB Abingdon, UK.
4) FOM institute DIFFER, Association EURATOM-FOM, P.O.Box 1207, 3430BE, Nieuwegein, Netherlands.
* annex to F. Romanelli et al, Proc. 24[rd] IAEA Conf., San Diego, 2012, IAEA Vienna.

**Abstract**
This paper will discuss simulations of the full ionization process (i.e. plasma burn-through), fundamental to creating high temperature plasma. By means of an applied electric field, the gas is partially ionized by the electron avalanche process. In order for the electron temperature to increase, the remaining neutrals need to be fully ionized in the plasma burn-through phase, as radiation is the main contribution to the electron power loss. The radiated power loss can be significantly affected by impurities resulting from interaction with the plasma facing components. The DYON code is a plasma burn-through simulator developed at Joint European Torus (JET) [1] [2]. The dynamic evolution of the plasma temperature and plasma densities including impurity content is calculated in a self-consistent way, using plasma wall interaction models. The recent installation of a beryllium wall at JET enabled validation of the plasma burn-through model in the presence of new, metallic plasma facing components. The simulation results of the plasma burn-through phase show consistent good agreement against experiments at JET, and explain differences observed during plasma initiation with the old carbon plasma facing components. In the International Thermonuclear Experimental Reactor (ITER), the allowable toroidal electric field is restricted to 0.35 [V/m], which is significantly lower compared to the typical value (~ 1 [V/m]) used in the present devices. The limitation on toroidal electric field also reduces the range of other operation parameters during plasma formation in ITER. Thus, predictive simulations of plasma burn-through in ITER using validated model is of crucial importance. This paper provides an overview of the DYON code and the validation, together with new predictive simulations for ITER using the DYON code.

**1. Introduction**

A gas discharge can be produced by applying an electric field to a gas. The seed electrons are accelerated and via collisions they ionize the neutrals producing more electrons and ions. These new electrons are again accelerated and make further impact ionizations producing an electron avalanche. Moreover, secondary electrons are generated at the cathode by impacting ions, which can generate further electrons through the electron avalanche process. This process maintains the discharge. The physics of gas discharge formation was first explained by Townsend, and is called a Townsend break-down [3].

At low temperatures (< 100 eV), the plasma is not yet fully ionized. Examples of partially ionized discharges can be found in laboratory plasmas as well as during the start-up phase of fusion devices. Neutrals and partially ionized ions emit line radiation which could result in the loss of a significant part of the (ohmic) heating power. The radiated power is proportional to the product of electron density and neutral density. As ionizations proceed, the electron density increases and the neutral density decreases, resulting in the maximum of radiated power at certain degree of ionization, which is called radiation barrier. If the heating power exceeds the radiation barrier, the radiated power loss decreases and the electron temperature can increase.

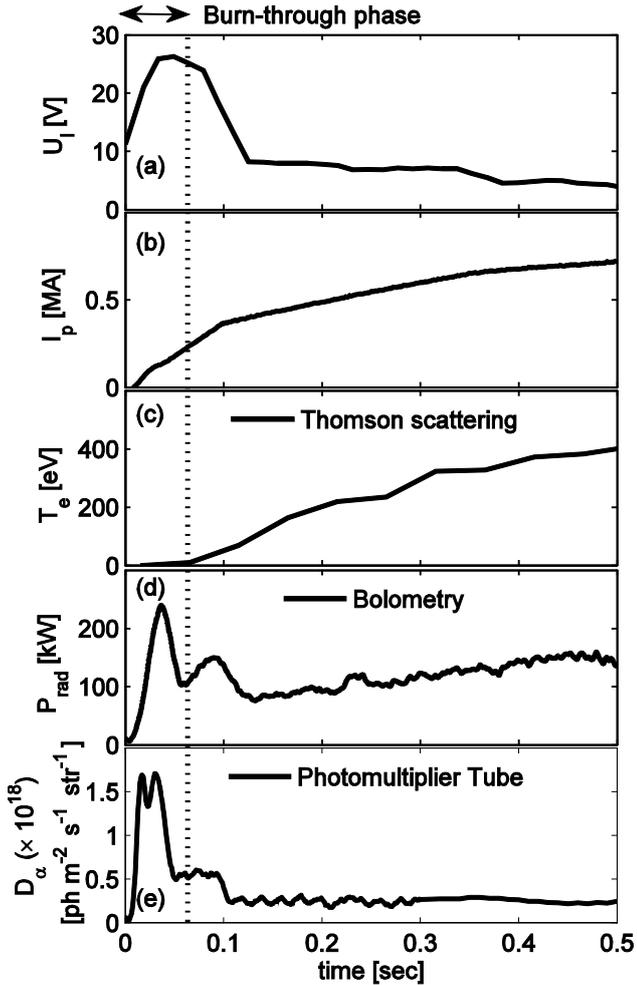

**Figure 1** Typical experimental data during the plasma burn-through phase measured in JET (#77210); (a) Toroidal loop voltage $U_l$ (Note, at JET $E_l \sim U_l/20m$), (b) Plasma current $I_p$, (c) Electron temperature $T_e$ (Thomson scattering), (d) Radiated power $P_{rad}$ (Bolometry), (e) D alpha emission (Photomultiplier tube).

In fusion research, the most promising device for magnetic confinement is the Tokamak, where the plasma is confined in a toroidal vacuum vessel and kept away from the solid walls by magnetic fields. The so-called effective connection length $L_f$ [4] is commonly used as a measure of the travelling distance to the surrounding wall; $L_f$ is the average length of the magnetic field lines between two contacting points at the vessel wall, and is in the range 100m ~ 1000m (without a plasma current). A toroidal electric field can be applied to a filling gas (typically deuterium, with pressure $p(0) \sim 5 \times 10^{-5}$ Torr) by the variation of the current in the central solenoid. In present devices the magnitude is of the order of 1 V/m. By this electric field, electron avalanche can occur in the prefill gas, and a plasma current is generated. (Note, for air at standard temperature and pressure, the electric field needed to generate arc between 1 meter gap electrodes is about 3.4 MV/m.) If ohmic heating of the plasma current is high enough to overcome the radiation barrier, the ionization process continues, thereby enabling $T_e$ to increase.

Full ionization of the prefill gas is called plasma burn-through. Figure 1 shows typical plasma burn-through phase in the Joint European Torus (JET). High toroidal loop voltage is applied at the beginning of discharge. The decrease in $P_{rad}$ and D alpha emission after the radiation barrier in Figure 1(d) and the emission peak in Figure 1(e) indicates that the prefill D atoms are ionized since both data are proportional to the D atom density.

The increase in electron temperature is very important for tokamak start-up. The plasma resistance decreases as electron temperature increases. Hence, the plasma current increases with electron temperature. After plasma burn-through the electron temperature keeps increasing even with much smaller loop voltage (See Figure 1 (a)). The build-up of plasma current generates poloidal magnetic field, which makes the magnetic field lines closed, thereby resulting in infinite effective connection length. When the effective connection length is not long enough, the parallel transport along the magnetic field lines is the dominant transport mechanism. Thus, the increase in plasma current improves the plasma confinement in a tokamak.

Plasma burn-through is determined by several parameters [5]: (1) the induced toroidal loop voltage $U_l$, (2) the prefill gas pressure $p(0)$, (3) the effective connection length $L_f$, (4) the initial impurity content of the discharge $n_I(0)$, and (5) the ratio of plasma volume $V_p$ to vessel volume $V_V$. Computational simulations are required to take into account the effects of these parameters.

In this paper, we review in section 2 the models used for simulating the evolution of plasma parameters during the plasma burn-through phase. Impurities in a plasma can result in significant radiated power loss, which would lead to failed plasma burn-through. The model includes a self-consistent simulation of the impurity content of the plasma by including plasma wall interaction. In section 3, the model used for plasma wall interaction is explained.

Previously, plasma burn-through simulations were attempted for International Thermonuclear Experimental Reactor (ITER) using constant impurity fraction [6] or simple exponential function of time [7] [8]. However, those simulations have never been compared against experimental data in present devices. The DYON code is a plasma burn-through simulator developed at JET. It has been validated using JET experiment results. In section 4, we review a validation of the simulations using JET data, which was published in [1] (Carbon wall) and [2] (Beryllium wall).

Succeeding work from the previous publication [2] is presented in the rest of the paper. Using the models and assumptions validated with the JET ITER-like wall, new predictive simulations for ITER are provided section 5 and 6. The plasma facing components of the ITER will be made of Be and W [9]. ITER is two times larger than JET, and the use of superconducting coils restricts the toroidal loop voltage available for plasma initiation. As a result, ITER will have a limitation on toroidal electric field of 0.35 V/m [8]. This is small compared to the typical toroidal electric field in JET, which is about 1 V/m. This limitation of toroidal electric field will significantly reduce the operation window for burn-through. With the design value of ITER, the requirement for successful ohmic burn-through will be addressed in section 5.

For reliable start-up in ITER, additional RF power is planned [10]. Additional RF power reduces the required toroidal E field since it provides seed electrons (i.e. pre-ionization) needed for electron avalanche [11]. Furthermore, the additional RF heating can assist plasma burn-through as shown in tokamak experiments such as AUG [10] and DIII-D [12] [13]. In section 6, the simulations of RF-assisted plasma burn-through in ITER are presented, and the required RF power is estimated. Finally, the DYON simulation results are compared with the previous predictions using the 0D code [6] and the SCENPLINT code [8].

## 2. Simulations of the plasma burn-through phase

The energy flow in a plasma is important to simulate the plasma burn-through phase (see Figure 2). Free electrons gain energy from ohmic heating or additional RF power, and lose the energy through three channels: (1) Transport: as free electrons lose their kinetic energy when they escape the plasma, (2) Radiation and Ionization: free electrons lose the energy when they collide with bound electrons in ions or atoms and (3) Equilibration: free electrons lose the energy by temperature equilibration with ions. Thus, equilibration is a heating channel for ions. The ions lose energy by transport and charge exchange.

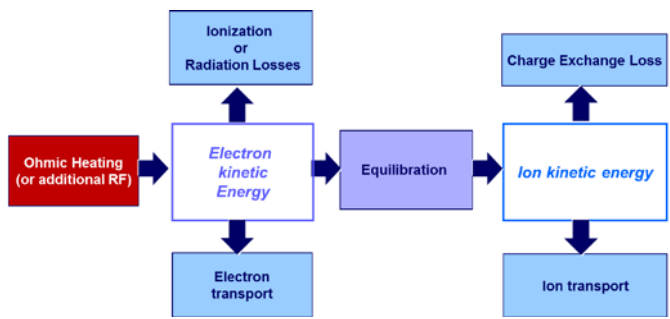

**Figure 2** Energy balance in a plasma during the plasma burn-through phase.

In order to solve this energy balance, one needs to solve the particle balance simultaneously (see Figure 3), since density terms are involved in the energy balance. A prefill gas of the vacuum chamber provides deuterium (D) atoms. The D atoms are ionized, and D ions recombine with free electrons to return to neutrals. Deuterium ions are also transported to the wall, resulting in recycling of D atoms or sputtering of impurity atoms.

Impurity atoms are ionized, and can recombine with free electrons reducing their charge state. Impurity ions are also transported to the wall; the impurity wall-sputtering generates additional impurity atoms. The impurity ions can accept an electron from other atoms or ions. In the simulations presented here, it is assumed that only D atoms are an electron donor in charge exchange reactions. All the atomic reactions are functions of plasma temperature, so energy balance should be solved simultaneously.

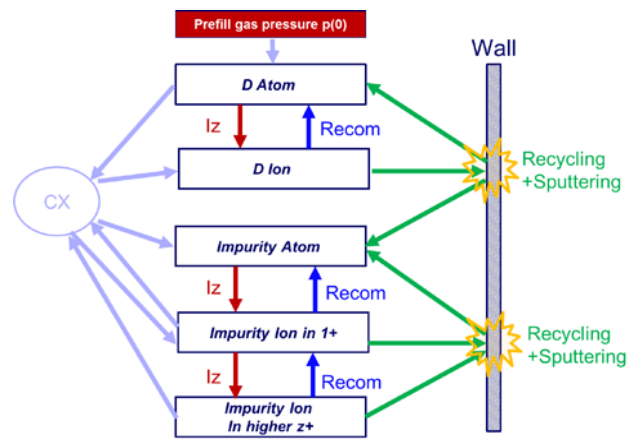

**Figure 3** Particle balance in a plasma during the plasma burn-through phase. CX is charge exchange reaction between D atoms and impurity ions, Iz is impact ionization by free electrons, Recom is Recombination reaction. Wall is the surrounding plasma facing component; in this study the wall material is carbon or beryllium.

The DYON code computes the energy balance equations and particle balance equations to determine the evolution of temperature and density in a plasma during the burn-through phase. Impurity densities are calculated for all charge states. Using the computed electron temperature, the plasma resistance is obtained, and a circuit equation calculates plasma current and the resultant ohmic heating. The ohmic heating is included in the energy balance. From this coupled differential equation system, the DYON code computes self-consistent values of the plasma parameters in time. More details on the DYON code can be found in [1].

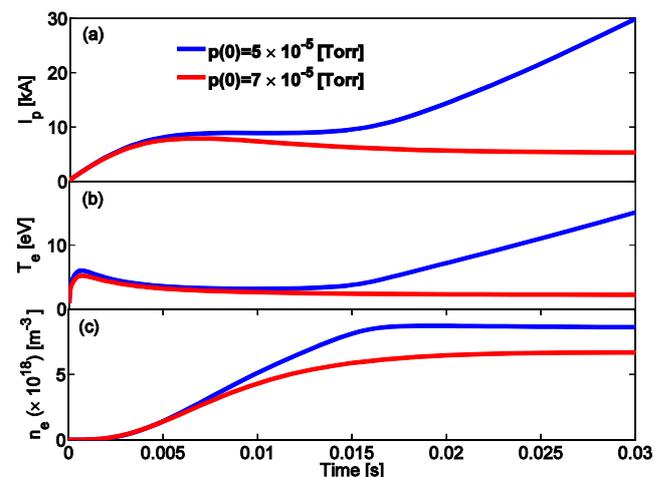

**Figure 4** Typical DYON simulation results for a pure D plasma in JET (R=3m and a=0.5m); (a) plasma current, (b) electron temperature, and (c) electron density. Under the same toroidal loop voltage 20 V, different prefill gas pressures are given; p(0)= 5x10$^{-5}$ Torr (blue) and 7x10$^{-5}$ Torr (red).

An example of a simulation with the DYON code for a pure D plasma is given in Figure 4. For the same loop voltage (20 V), but for two different values of the prefill gas pressures (5x10$^{-5}$ Torr and 7x10$^{-5}$ Torr), the results show a successful and failed case of plasma burn-through in JET. In Figure 4, the blue and red lines are the simulation results at low and high prefill gas pressures, respectively. At the low prefill gas pressure, full ionization is obtained, and the electron temperature increases together with the plasma current. In case of a high prefill gas pressure, the electron density is lower than that at the low prefill gas pressure as the plasma is not fully ionized. Due to the significant radiated power loss, the electron temperature and plasma current saturate to constant values. Since a constant loop voltage is used in the simulations, the plasma current does not collapse even in the case the burn-through fails. In contrast, in the JET experiments the toroidal loop voltage is preprogrammed to be reduced after 100 ms. Thus, the plasma current is not maintained in case plasma burn-through is unsuccessful.

## 3. The inclusion of plasma wall interactions in the simulations

Impurity effects are essential for computing the radiated power. Most of impurities come from the surrounding wall via Plasma Wall Interactions (PWI) such as sputtering or recycling. A new all metal wall (called the ITER-like wall) has been installed recently in JET, made of Beryllium (Be) in the main chamber (e.g. important for break-down physics) and Tungsten (W) tiles in the lower part of the vessel called divertor [14]. Experiments show a clear difference in the measured radiated power during the plasma burn-through phase with the different wall materials in JET; previous experiments with a Carbon (C) wall have much higher radiated power than the Be wall [15]. The magnitude of radiation barrier was found to be linearly correlated with the calculated impurity influx from the C wall, while such a relation was not found for the Be ITER-like wall [16].

A PWI model has been included in the DYON code (see Figure 3.) [1] [2]. The ion flux to the wall $\Gamma_{out}^{z+}$ can be calculated using the plasma volume $V_p$, the ion density $n_{out}^{z+}$ and the particle confinement time $\tau_p$;

$$\Gamma_{out}^{z+} = V_p \frac{n_{out}^{z+}}{\tau_p} \quad (1)$$

The ion outflux results in D recycling at the wall which is calculated using the recycling coefficient $Y_D^D$. Based on experimental data, $Y_D^D$ can be adjusted. Previous simulations use a recycling coefficient (> 1) during plasma formation for the C wall, decaying to 1, while for the ITER-like wall a recycling coefficient (< 1) was used, increasing towards 1 [2]. The ion outflux also results in impurity sputtering at the wall. Impurity sputtering can be calculated by using the corresponding sputtering yield $Y_{in}^{out}$. The resultant neutral influx to a plasma $\Gamma_{in}^0$ is computed using:

$$\Gamma_{in}^0 = \sum_{out} \sum_{z \geq 1} Y_{in}^{out} \times \Gamma_{out}^{z+} \quad (2)$$

In the case of a Be wall, physical sputtering is dominant due to the low threshold energy [17]. In the case of a carbon wall, chemical sputtering is dominant, resulting in production of hydrocarbon and carbon monoxide at all impact energies [18].

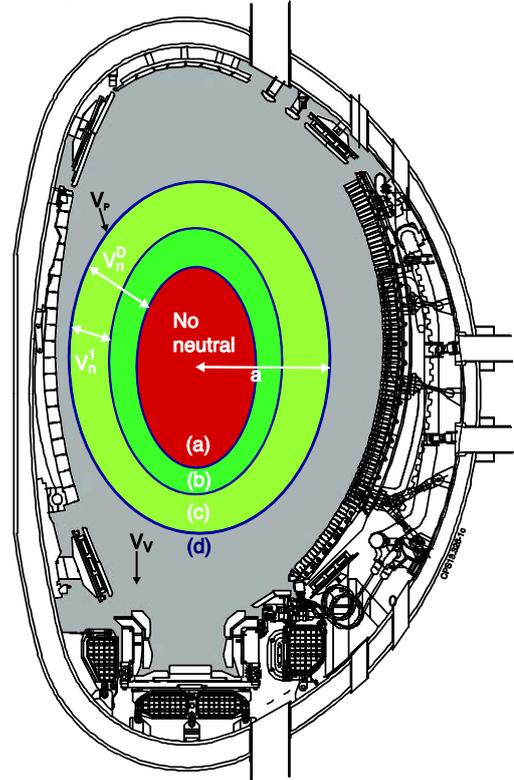

**Figure 5** Ionization shell model. $V_p$ (= (a) + (b) + (c)) is a plasma volume, and $V_V$ (=(a)+(b)+(c)+(d)) is the total vessel volume. Since there are no neutrals in (a), the total neutral volume is (b)+(c)+(d). $V_n^D$ and $V_n^I$ are the neutral volumes of Deuterium (=(b)+(c)) and Impurity(=(c)) within the plasma volume $V_p$, respectively.

From the plasma wall interactions, D and impurities enter the plasma, and are ionized in the outer part (shell) of the plasma volume (see Figure 5). The shell volume is determined by the mean-free-path for ionization. During the burn-through phase, the mean free path decreases as the electron temperature increases. If the mean-free-path is longer than the plasma size (minor radius), the volume occupied by neutrals within the plasma is equal to the whole plasma volume. If it is shorter than the minor radius, neutrals will be absent in the core. More detailed calculations are available in [1]. The neutral screening effect determines the neutral volume within the plasma, which is important when calculating atomic reactions [6]. As a result, the volume where D atoms exist is somewhat larger than that of impurity atoms.

Impurity densities in all charge states should be calculated to determine the radiated power. Since the densities of the different charge states are dependent on each other, they should be solved in a matrix, also taking into account the impurity influx from the wall and the neutral screening effects. In addition to impurities entering the plasma via PWI, a (small) initial impurity fraction $n_I(0)$ in the residual gas might be present. A possible source could be those impurities only weakly attached to the wall after migration during the previous experiments. Starting with these initial impurity fractions, the impurity content during the burn-through phase is calculated self-consistently by the DYON code, using the plasma wall interaction models.

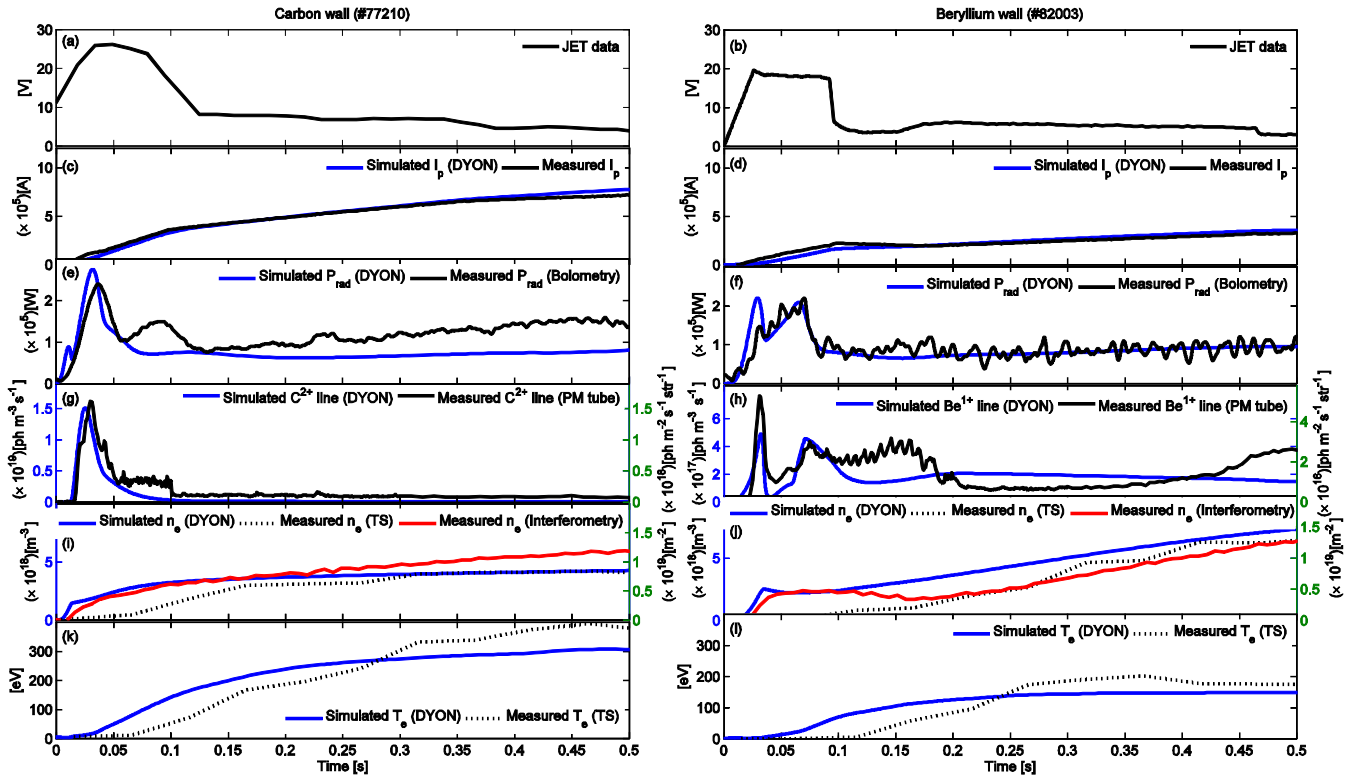

**Figure 6** Comparison of DYON simulation results and JET experiments with the C wall (#77210) **[1]** and the Be wall (#82003) **[2]**. (a)(b) toroidal loop voltage, (c)(d) plasma current, (e)(f) total radiation power, (g)(h) photon emission (C2+ and Be1+), (i)(j) electron density, (k)(l) electron temperature. (TS is Thomson Scattering.)

## 4. Validation of the simulations with experimental data from JET

The simulation results have been compared with the measured examples of plasma burn-through obtained in JET experiments with the C wall and the Be wall. Input such as toroidal loop voltage $U_l$, prefill gas pressure $p(0)$, magnetic fields ($B_t$ and $B_v$ to calculate $L_f$), and plasma size (R, a, and $V_p$) are obtained from measurement in JET. Different PWI models are adopted for the C wall and the Be wall. For the C wall, chemical sputtering is assumed to be dominant. For the Be wall, physical sputtering model is used. Dynamic D recycling, additional fuelling, and initial impurity contents are also modelled according to the different wall. Since not all neutrals in the vessel are accessible to the plasma, effective vessel volume is given as 100 m$^3$ (actual JET vessel volume is 189 m$^3$).

The validation of the DYON code using JET data with the C wall and the Be ITER-like wall has been published, and more details can be seen in [1] and [2], respectively. Here, we review the main validation results. With the given toroidal loop voltage scenario in Figure 6(a)(b), the plasma current is simulated showing very good agreement with the measurements. (see Figure 6(c)(d)) The simulated synthetic photon emission data for the impurities from the wall (Figure 6(g) C2+ and Figure 6(h) Be1+) have the same temporal evolution as those measured by photomultiplier tubes. Also, the measured total radiation power and temporal evolution are matched well by the DYON simulations in Figure 6(e)(f). The good agreement between the simulation results and the experimental data implies that important physics aspects of plasma burn-through, and also the dynamics thereof, are well modeled in the simulations.

The discrepancy of the simulated $T_e$ from the measured $T_e$ in Figure 6(k)(l) is probably due to the large error bar in the Thomson scattering data in the early phase, where the measured signal is very weak. The good agreement in plasma current in Figure 6(a)(b) implies that the electron temperature, which determines the plasma resistance, is properly calculated. Simulated $n_e$ shows similar trend with the interferometry data, although the Thomson scattering data shows discrepancy due to the same reason mentioned. However, both simulated $n_e$ with DYON and measured $n_e$ with Thomson scattering have similar values near 0.5 seconds.

The DYON code is used to compute and document the differences between a C wall and a Be wall. For the C wall, the radiation barrier is much higher and dominated by C radiation. However, in the Be wall, the radiation power loss is much smaller and not dominated by the Be radiation; The radiation barrier in the Be wall is dominated by the D radiation as long as other impurities are not significant [16].

Using the DYON code, the operation space for JET start-up can be determined. The required electric field for a range of prefill pressures has been computed [2]. The Townsend criterion provides a first estimate for the required electric field for electron avalanche, which is given by

$$E\,[V/m] = \frac{1.25 \times 10^4 \, p[Torr]}{\ln(510 \times p[Torr] \times L_f[m])} \quad . \quad (3)$$

In Figure 7, the required electric field for an effective connection length of 200m and 500m is shown.

The DYON simulations provide the required electric field for a successful plasma burn-through for a pure D, in the presence of either a Be or C wall. The resultant operation spaces are compared in Figure 7. The black solid line is the required toroidal electric field in pure D. At the range of prefill gas pressure above 5 x $10^{-5}$ Torr, the required electric fields for plasma burn-through (i.e. burn-through criterion) are much higher than the Townsend criterion alone.

It was observed that the Be wall tends to absorb the fuel at the wall. This can reduce the prefill gas pressure or initial plasma density, thereby resulting in slow electric avalanche (which is not favorable for operation) or run-away electron generation. Hence, the prefill gas pressure used for start-up with the Be wall in JET is higher than that of the C wall as shown in Figure 7.

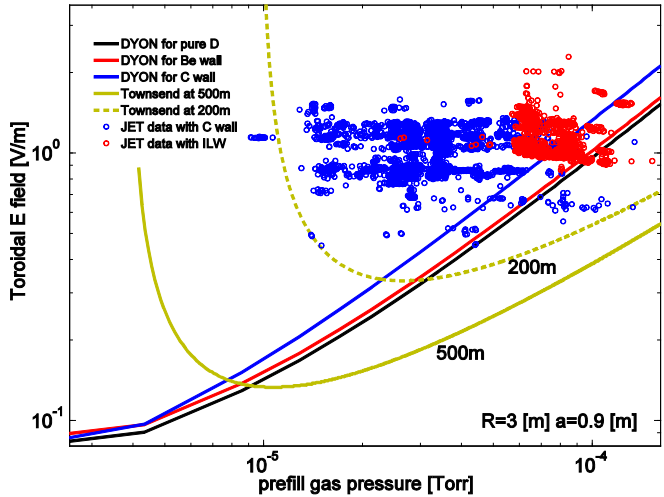

**Figure 7** Operation space for JET start-up. The green lines indicate the Townsend criterion for electron avalanche with an effective connection length of 200m (dashed) and 500m (solid), respectively. The black line is the required electric field for pure deuterium burn-through, computed using the DYON code. The red (Be wall) and blue (C wall) lines are the corresponding burn-through criterion including the different wall models. The circles correspond the successful plasma burn-through in JET experiments with the ITER-like wall (#80239 ~ #82905, red circles) and the C wall (#70988 ~ #78805, blue circles).

With the PWI models, the burn-through criterions increase from that for pure D plasma. While slightly more toroidal electric field is required with a Be wall, the burn-through criterion significantly increases in the presence of a C wall. These predictions are compared to experimental data using more JET data i.e. the ITER-like wall (#80239 ~ #82905, red circles) and the C wall (#70988 ~ #78805, blue circles). The majority of the experimental data points with the C wall (blue circles) and the Be wall (red circles) lie above (or near) the simulated minimum electric field required for a C wall (blue line) and a Be wall (red line) again showing good agreement of the simulations with the experiments, over a wide range on conditions. It should be noted that in previous publications, operation space for tokamak start-up was only calculated using the Townsend criterion [19]. However, Figure 7 shows that at high prefill gas pressure the limitation is set by the burn-through criterion rather than the Townsend criterion.

## 5. Ohmic plasma burn-through for ITER

| Plasma parameters | Input value |
|---|---|
| $U_l$ [V] | 11.36 at t=0.909[s] (see Figure 8 (a)) |
| $P_{RF}$ [MW] | 0 (Figure 7 and 8) <br> 4 (Figure 10) |
| $B_V$ [Tesla] | $4.4 \times 10^{-4}$ at t=0.909[s] |
| $B_R$ [Tesla] | $5.1 \times 10^{-5}$ at t=0.909[s] |
| R [m] | 5.65 |
| a [m] | 1.6 |
| $V_V$ [m$^3$] | 1400 |
| $B_\phi$ [Tesla] | 5.3 |
| p(0) [Torr] | $7.2 \times 10^{-6}$ (Figure 7 and 8) <br> $1.8 \times 10^{-5}$ (Figure 7) <br> $5 \times 10^{-5}$ (Figure 9 and 10) |
| $n_D^0(0) [m^{-3}]$ | $2.78 \times 10^{23} \times p(0)[Torr]$ |
| $n_{Be}^0(0) [m^{-3}]$ | $0.01 \times n_D^0(0)[m^{-3}]$ |
| $n_C^0(0) [m^{-3}]$ | $0.005 \times n_D^0(0)[m^{-3}]$ |
| $n_O^0(0) [m^{-3}]$ | $0.001 \times n_D^0(0)[m^{-3}]$ |
| $T_e(0)$ [eV] | 1 |
| $T_i(0)$ [eV] | 0.03 |
| $J_p(0) [Am^{-2}]$ | $382.5 \times E$ [V/m] |
| $Y_D^D$ | 1 |
| $\gamma_{iz}(0)$ | 0.002 |
| $l_i$ | 0.5 |

**Table 1** Input values given to the DYON code for plasma burn-through simulations of ITER

Plasma burn-through simulations for ITER have been performed with a PWI model for a Be wall, including the physical sputtering model as validated with the ITER-like wall in JET. The design values for ITER break-down obtained from F4E[1] are used in DYON simulations; i.e. the toroidal loop voltage $U_l(t)$, the vertical magnetic field $B_V(t)$ and radial magnetic field $B_R(t)$ which are used to compute the effective connection length, the plasma major $R(t)$ and minor radius $a(t)$, and the vessel volume $V_V$. An overview of the parameters is given in Table 1. In these simulations of plasma burn-through in ITER initial concentrations of Be (1% of $n_D(0)$), C (0.5% of $n_D(0)$), and O(0.1% of $n_D(0)$) are used. Other initial conditions and assumptions e.g. $T_e(0)$, $T_i(0)$, $J_p(0)$, $Y_D^D$, $l_i$ are also listed in Table 1.

The design value of the toroidal loop voltage reaches a maximum on axis at about 0.9s due to the shielding effect of the external magnetic fields at the vacuum vessel in ITER [8]. The plasma break-down (electron avalanche) is assumed to occur at 0.9s, so the plasma burn-through simulations start at t=0.9s. The toroidal loop voltage is preprogrammed to decrease after 0.9s (see Figure 11 (a)).

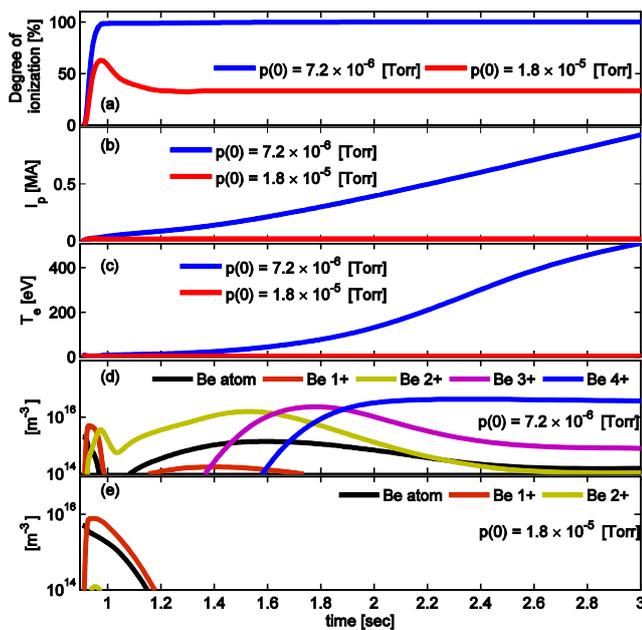

**Figure 8** Predictive ohmic plasma burn-through simulations for ITER. Using design values ($V_L(t)$, $B_v(t)$, $B_R(t)$, $R(t)$, and $a(t)$), different prefill gas pressures are used; p(0)= 7.2 x 10$^{-6}$ Torr (Blue) and 1.8 X 10$^{-5}$ Torr (Red) in (a) Degree of ionization, (b) Plasma current, and (c) Electron temperature. (d) Evolution of impurity density (Be) in ITER simulations at p(0)= 7.2 x 10$^{-6}$ Torr, and (e) p(0)=1.8 X 10$^{-5}$ Torr. Be density in each charge state in (d) and (e) are indicated by different color.

With the given design values, it is found that very low prefill gas pressures (7.2 x 10$^{-6}$ Torr) are required to achieve ohmic plasma burn-through (see Figure 8 (a)). Higher prefill gas pressures (1.8 x 10$^{-5}$ Torr) used in present day devices (typical p(0) in JET is 5 x 10$^{-5}$ Torr) will result in ITER in a failed burn-through, despite ITER having a connection length of $L_f$=4000 m at the start of the simulation (t = 0.9s). The required time to achieve 1 MA of plasma current is about 2 seconds (see Figure 8 (b)), which is much longer compared to that in JET (0.5 ~ 1s).

Figure 8 (d) and (e) show the corresponding Be evolution in the ITER simulations. The densities for the different charge states of Be are computed in the DYON code. In the case of a successful plasma burn-through, Be$^{4+}$ becomes the dominant charge state. However, in the failed case, the initial Be content does not continue the ionization process and the total content decreases. If plasma burn-through is not successful, the plasma temperature does not increase. This results in the incident ion energy below the threshold energy for physical sputtering i.e. no further source of Be.

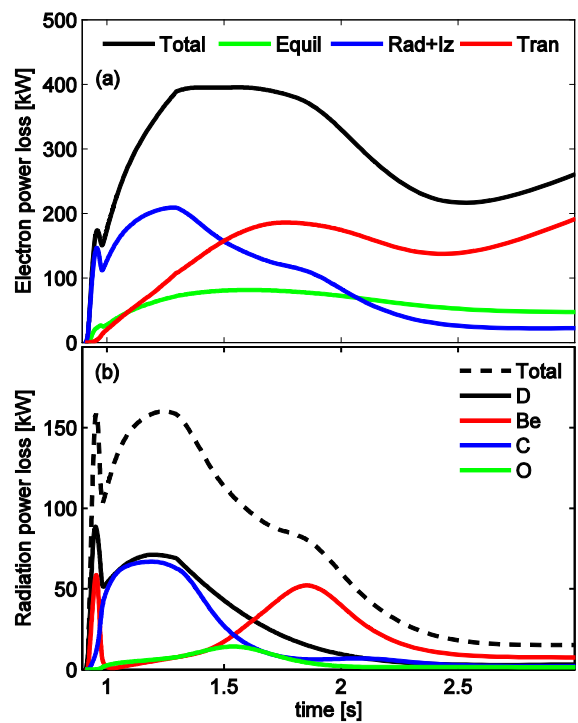

**Figure 9** Simulated power losses for the simulation of ohmic plasma burn-through at p(0)= 7.2 x 10$^{-6}$ Torr in Figure 8. (a) electron power losses; Total electron power loss (black), Electron power loss due to the radiation and ionization (blue), Transport power loss (red), and Equilibration power loss (green). (b) radiation power loss; The total radiated power is indicated by the black dashed line. The contribution to the radiated power are indicated; D (black), Be (red), C (blue), and O (green).

Figure 9 shows (a) the electron power loss and (b) the radiation power loss for the successful simulation of plasma burn-through in ITER. The electron power loss during the early start-up phase up to t=1.5s is dominated by the radiation and ionization power losses (blue in Figure 9(a)). However, towards the end of the plasma burn-through process, the power loss due to the electron transport dominates. Figure 9(b) shows the corresponding radiated power. The sources of the radiation are shown by different colors. The majority of the radiated power during the plasma burn-through phase is D radiation (black line). The first peak in Be radiation (red line) at the start of the discharge, results from the initial Be content. It decreases immediately after the initial radiation peak. The second increase in the Be radiation around t=1.9s is due to more Be entering the plasma due to physical sputtering. Until the incident ions have reached sufficient temperature (i.e. the incident energy exceeds the physical sputtering threshold) this second Be radiation peak is delayed. This evolution (double peak) of the Be radiation has been observed in experiments with the ITER-Like Wall in JET. Thus, Be radiation is not likely to impact on plasma burn-through in ITER. The initial C content used in the simulations is based on the JET data with the ITER-like wall [20]. This indicates that the C radiation is comparable to the D radiation for the period t=1 ~ 1.5s (see Figure 9(b)). For a higher initial C content the C radiation will dominate the radiation barrier.

---
[1] http://www.fusionforenergy.europa.eu/

## 6. RF-assisted plasma burn-through for ITER

RF power can provide pre-ionization for electron avalanche and additional heating during the plasma burn-through phase. Thus, in ITER, RF power will be used to ensure robust start-up. Using the DYON code, RF-assisted plasma burn-through has been simulated for ITER. Here, the RF power is defined as the absorbed RF power in the plasma. If the given RF power is high enough for full ionization of the plasma and to overcome the radiation barrier, the plasma current increases. Figure 10 shows the DYON simulation results. There is a critical RF power for plasma burn-through between 3MW and 4MW at prefill pressure of 5x$10^{-5}$ Torr (with initial conditions given in Table 1).

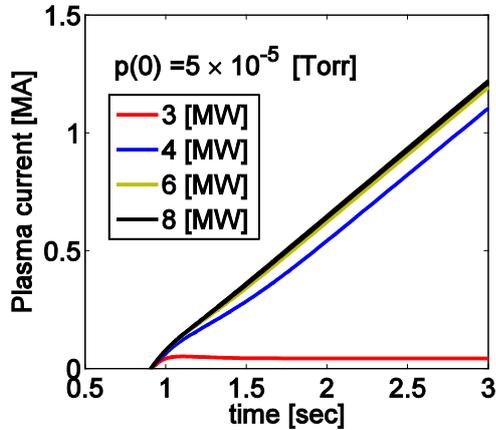

**Figure 10** Simulated plasma current in RF-assisted plasma burn-through simulations for ITER. With the same prefill gas pressure (5 x $10^{-5}$ Torr), the RF power is scanned.

Once plasma burn-through is completed, the ramp-up rate of plasma current is mainly determined by the toroidal loop voltage. The plasma current keeps increasing even if the RF power is switched off (e.g. the RF power could be turned off at t=2.55 s). In addition, the plasma current ramp-up does not increase much faster with higher RF power (6 MW (green) and 8 MW (black) in Figure 10).

This can be explained by the circuit equation of plasma current below.

$$\frac{dI_p}{dt} = \frac{U_l}{L_p} - \frac{I_p R_p}{L_p} \quad (3)$$

where $U_l$ and $L_p$ are toroidal loop voltage and self-inductance of plasma current, respectively. It is assumed in the simulations that both are constant during the $I_p$ ramp-up phase. Hence, the maximum of the $I_p$ ramp-up rate is set by $\frac{U_l}{L_p}$. The slightly faster increase in plasma current (4MW (blue) and 6MW (green) in Figure 10) is due to the faster increase in electron temperature, thereby reducing the plasma resistance.

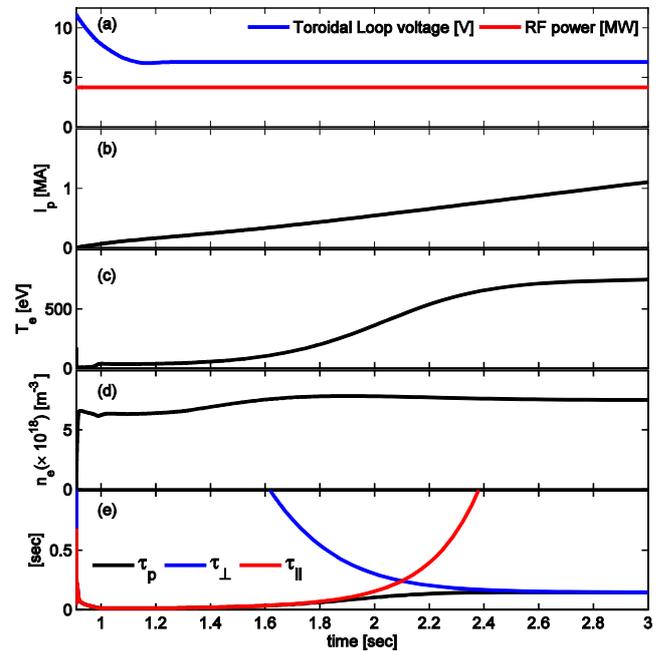

**Figure 11** Simulation results of RF-assisted plasma burn-through in ITER. (a) Toroidal loop voltage (design value obtained from F4E. Note, at ITER $E_l \sim U_l/40m$) and RF power (4MW, which is the minimum RF power required for 5 x $10^{-5}$ Torr of prefill gas), (b) Plasma current, (c) Electron temperature, (d) Electron density, and (e) Particle confinement time; perpendicular confinement time (blue), parallel confinement time (red), and the resultant particle confinement time (black)

Figure 11 shows a DYON simulation of RF-assisted plasma burn-through in ITER. In this simulation, constant 4MW of RF power and the same design values in Table 1, including the toroidal loop voltage (see Figure 11 (a)), are used. Although the prefill gas pressure used (5x$10^{-5}$ Torr) is much higher than that for ohmic plasma burn-through (7.2x$10^{-6}$ Torr) in Figure 8, the plasma current and electron temperature are slightly higher in the simulations of RF-assisted burn-through. Deuterium burn-through is completed instantly as can be seen by the abrupt increase in electron density in Figure 11(d).

Figure 11(e) shows the particle confinement time $\tau_p$. In the DYON simulations, the particle confinement time $\tau_p$ is computed as

$$\frac{1}{\tau_p} = \frac{1}{\tau_\parallel} + \frac{1}{\tau_\perp} \quad (4)$$

where $\tau_\parallel$ is parallel confinement time obtained by transonic ambipolar transport, and $\tau_\perp$ is perpendicular confinement time calculated by Bohm diffusion model [1]. At the beginning of the plasma burn-through phase, where the effective connection length is not sufficiently long, particle transport is dominated by the parallel transport along the magnetic field lines. As the plasma current increases, closed magnetic flux surfaces are formed; the effective connection length approaches infinity. This reduces the parallel transport, and the perpendicular transport becomes dominant (See Figure 11 (e)).

The key features of the previous plasma burn-through simulators and the DYON code are summarized in Table 2. The absorbed power given in the overview table means the value for reliable plasma burn-through given in the publications, rather than the minimum required RF power. The required RF power is subject to operation parameters such as prefill gas pressure and initial impurity content. Figure 12 gives the DYON estimation of the required RF power in ITER at various operation parameters showing the impact of prefill gas pressure p(0) and initial impurity fraction $n_I(0)$. In Figure 12(a) a scan of initial C fraction is given and a scan of initial Be fraction is given in (b). Fixed initial Be fraction (1% of $n_D(0)$) and initial C fraction (1% of $n_D(0)$) are used in (a) and (b), respectively. As shown by Figure 12 (a) and (b), the initial Be fraction does not impact on the required RF power, but the initial C content results in significant differences in the required RF power. In both cases, without RF assist, plasma burn-through will be possible only at prefill gas pressure below 1 x $10^{-5}$ Torr. The required RF power increases almost linearly with the prefill gas pressure. This is due to an increase in D atom density with prefill gas pressure. In addition, since we assume here the fraction of the initial impurity against prefill D density, the impurity content is also higher with higher D prefill gas pressure.

It is planned to have RF power up to 8MW available in ITER for burn-through assist. If the initial C content is smaller than the assumed 1% of prefill D atoms $n_D(0)$, using the RF assist power, plasma burn-through in ITER will be available at around 5 x $10^{-5}$ Torr, which is a typical prefill gas pressure used in present devices including JET.

## 7. Discussions and conclusions

A plasma burn-through simulator, the DYON code, has been developed for partially ionized plasmas at low electron temperature (in the burn-through phase). Solving the differential equation system of energy balance, particle balance, and circuit equations, the DYON code computes the evolution of plasma parameters such as $I_p$, $T_e$, $n_e$, and $n_I$ during the plasma burn-through phase. During the plasma burn-through phase, the impurities resulting from the wall can increase the radiated power loss significantly, thereby causing the failure of plasma burn-through. Hence, the impurity evolution is computed self-consistently using a model for plasma wall interaction.

| | 0D code [6] | SCENPLINT [8] | DYON [1][2] |
|---|---|---|---|
| Impurity model | constant fraction $\frac{n_{Be}(t)}{n_D(0)} = 0.05$ or $\frac{n_C(t)}{n_D(0)} = 0.05$ | exponential growing fraction $\frac{n_{Be}(t)}{n_D(0)} = 0.02(1-\exp(\frac{-t}{0.25}))$ or $\frac{n_C(t)}{n_D(0)} = 0.013 + 0.03(1-\exp(\frac{-t}{0.25}))$ | Plasma wall interaction $\Gamma_{in}^0 = \sum_{out}\sum_{z\geq 1} Y_{in}^{out} \times \Gamma_{out}^{z+}$ where $\Gamma_{out}^{z+} = V_p \frac{n_{out}^{z+}}{\tau_p}$ |
| Transport model | constant confinement time or INTOR scaling law $\tau_p$ = constant (5 ~ 50 ms) $\tau_E = 5\times 10^{-21} n_e \tau_0$ where $\tau_0$ = constant (0.2~2) $\tau_{imp}$ = constant (20~100 ms) | Scaling law $\tau_p = \tau_E$ $\tau_E = \max(\tau_{Bohm}^{scaling}, \tau_{L-mode}^{ITER-98})$ where $\tau_{Bohm}^{scaling} = 3\times 10^{-3} a^2 B_t T_e$ $\tau_{L-mode}^{ITER-98}$ = ITER-98 L-mode confinement scaling | Transonic parallel transport + Bohm diffusion $\frac{1}{\tau_p} = \frac{1}{\tau_\parallel} + \frac{1}{\tau_\perp}$ $\tau_\parallel = \frac{L_f(t)}{C_s(t)}$ $\tau_\perp$ = Bohm diffusion $\tau_E = \tau_p$ |
| Absorbed RF power and prefill gas pressure for successful plasma burn-through | $P_{RF}$ = 2 ~ 5 MW p(0) = 2~7×$10^{-6}$ Torr | $P_{RF}$ = 2 MW p(0) = 6×$10^{-6}$ Torr | See Figure 12 |
| Validation with experiments | X | X | Compared to JET experiments with the ITER-like wall and the carbon wall |

**Table 2** Comparison of the simulators used to predict plasma burn-through in ITER

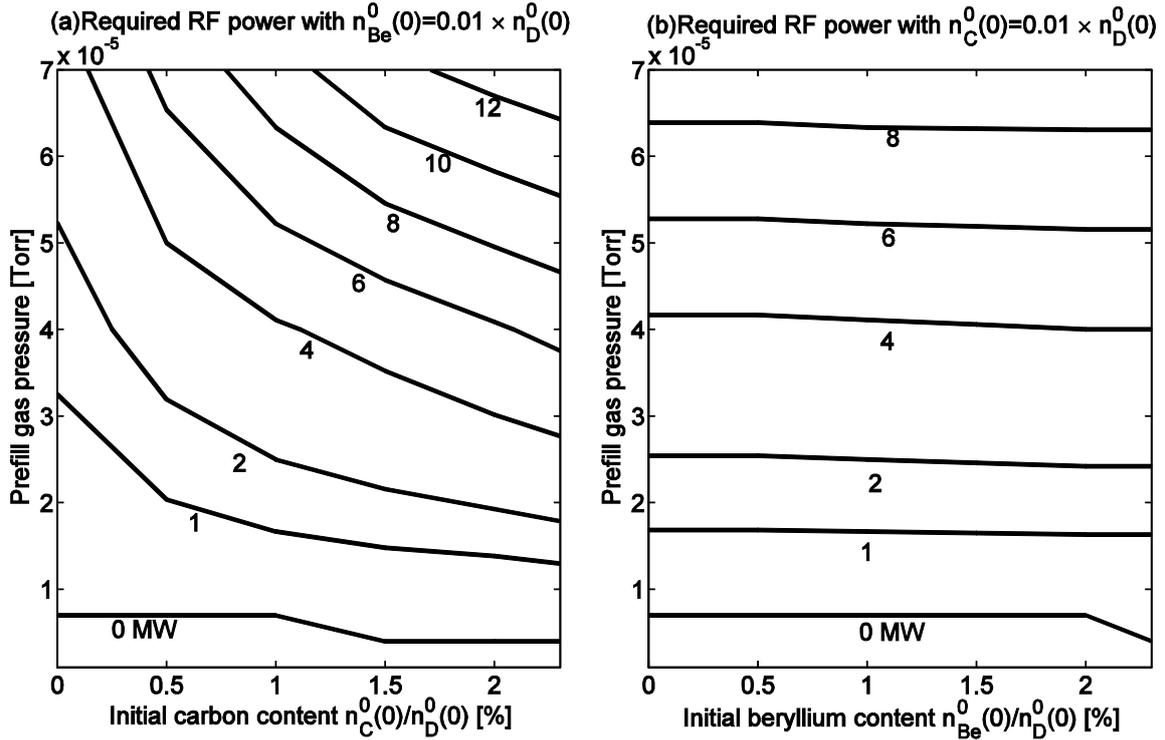

**Figure 12** Estimation of the required RF power using the DYON simulations for successful plasma burn-through in ITER. (a) initial Be content 1% of $n_D(0)$  (b) initial C content 1% of $n_D(0)$.

The simulation results have been compared with experimental data of JET ITER-like wall. The simulated diagnostic data such as total radiated power and photon emission show good agreement with the measured values in JET. Plasma current is reproduced matching very well with the JET data, and other simulated plasma parameters (i.e. $T_e$, and $n_e$) agree with experimental observations. Previously, the Townsend criterion was used to estimate the required operation space for tokamak start-up. The required operation space for plasma burn-through in JET has been computed using the DYON code. For the same toroidal loop voltage, the maximum prefill gas pressure available for plasma burn-through is much lower than that calculated by Townsend criterion. This implies that operation space is reduced by the burn-through criterion (even without impurities), so it should be taken account for devices in the future such as ITER.

The DYON code has been used to perform a predictive simulation for ITER. It should be noted that the prediction for ITER is based on the assumptions given in Table 1. Although the assumptions are selected according to the experimental data at JET ITER-like wall and design value from F4E, the required operation parameters can be different according to other assumptions e.g. a smaller plasma minor radius or effective vessel volume.

With the conditions given in Table 1, DYON simulations show that ohmic plasma burn-through in ITER (without RF assist) will be available only at very low prefill gas pressure (at $p(0) < 10^{-5}$ Torr). At JET with the ITER-like wall such lower prefill gas pressures are not used to avoid having a too slow electron avalanche phase or too low initial plasma density that may cause run-away electrons. However, 4MW of RF assist will make ITER start-up available at prefill gas pressures up to $5 \times 10^{-5}$ Torr, which is in the typical range used at present devices. The RF assist will result in instant deuterium burn-through. Once plasma burn-through is completed, the ramp-up rate of plasma current is not significantly affected by increasing the RF power.

The required RF power is subject to prefill gas pressure and initial impurity content. With prefill gas pressure, the required RF power increases almost linearly. The initial Be content will not impact on plasma burn-through in ITER, but an initial C content can increase the radiated power significantly. Fortunately, in case a full W diverter is installed in ITER, the initial C content in ITER is expected to be much lower than observed in JET experiments with the ITER-like wall  (0.5 ~ 1 % of $n_D(0)$).

**Acknowledgement**

This research was funded partly by the Kwanjeong Educational Foundation and by the European Communities under the contract of Association between EURATOM and CCFE. The authors thank F4E for providing the basic ITER parameters. The views and opinions expressed herein do not necessarily reflect those of the European Commission. This work was carried out within the framework of the European Fusion Development Agreement.